\documentclass[12pt,preprint]{aastex}

\newcommand\ReinL{R_{\rm Ein,L}}
\newcommand\ReinS{R_{\rm Ein,S}}

\received{}
\accepted{}

\slugcomment{12 February 2004; submitted to ApJ}

\author{
Gordon T. Richards,\altaffilmark{1}
Charles R. Keeton,\altaffilmark{2,3}
Bartosz Pindor,\altaffilmark{1,4}
Joseph F. Hennawi,\altaffilmark{1}
Patrick B. Hall,\altaffilmark{1}
Edwin L. Turner,\altaffilmark{1}
Naohisa Inada,\altaffilmark{5}
Masamune Oguri,\altaffilmark{5}
Shin-Ichi Ichikawa,\altaffilmark{6}
Robert H. Becker,\altaffilmark{7,8}
Michael D. Gregg,\altaffilmark{7,8}
Richard L. White,\altaffilmark{9}
J. Stuart B. Wyithe,\altaffilmark{1,10}
Donald P. Schneider,\altaffilmark{11}
David E. Johnston,\altaffilmark{1,2,12}
Joshua A. Frieman,\altaffilmark{2,12,13}
and J. Brinkmann\altaffilmark{14}
}

\altaffiltext{1}{Princeton University Observatory, Peyton Hall, Princeton, NJ 08544.}
\altaffiltext{2}{Department of Astronomy and Astrophysics, The University of Chicago, 5640 South Ellis Avenue, Chicago, IL 60637.}
\altaffiltext{3}{Hubble Fellow.}
\altaffiltext{4}{Canadian Institute for Theoretical Astrophysics, University of Toronto, 60 St.\ George Street, Toronto, Ontario, M5S 3H8, Canada.}
\altaffiltext{5}{Department of Physics, University of Tokyo, 7-3-1 Hongo, Bunkyo, Tokyo 113-0033, Japan.}
\altaffiltext{6}{National Astronomical Observatory, 2-21-1 Osawa, Mitaka, Tokyo 181-8588, Japan.}
\altaffiltext{7}{Physics Department, University of California, Davis, CA 95616.}
\altaffiltext{8}{IGPP-LLNL, L-413, 7000 East Avenue, Livermore, CA 94550.}
\altaffiltext{9}{Space Telescope Science Institute, 3700 San Martin Drive, Baltimore, MD 21218.}
\altaffiltext{10}{University of Melbourne, Parkville, VIC 3010, Australia.}
\altaffiltext{11}{Department of Astronomy and Astrophysics, The Pennsylvania State University, 525 Davey Laboratory, University Park, PA 16802.}
\altaffiltext{12}{Center for Cosmological Physics, The University of Chicago, 5640 South Ellis Avenue Chicago, IL 60637.}
\altaffiltext{13}{Fermi National Accelerator Laboratory, P.O. Box 500, Batavia, IL 60510.}
\altaffiltext{14}{Apache Point Observatory, P.O. Box 59, Sunspot, NM 88349.}

\begin{document}

\title{Microlensing of the Broad Emission Line Region in the
Quadruple Lens SDSS~J1004+4112}

\begin{abstract}

We present seven epochs of spectroscopy on the quadruply imaged quasar
SDSS~J1004+4112, spanning observed-frame time delays from 1 to 322
days.  The spectra reveal differences in the emission lines between
the lensed images.  Specifically, component A showed a strong
enhancement in the blue wings of several high-ionization lines
relative to component B, which lasted at least 28 days (observed
frame) then faded.  Since the predicted time delay between A and B is
$\lesssim$30 days, our time coverage suggests that the event was not
intrinsic to the quasar.  We attribute these variations to
microlensing of part of the broad emission line region of the quasar,
apparently resolving structure in the source plane on a scale of
$\sim\!10^{16}\,{\rm cm}$ at $z=1.734$.  In addition, we observed
smaller differences in the emission line profiles between components A
and B that persisted throughout the time span, which may also be due
to microlensing or millilensing.  Further spectroscopic monitoring of
this system holds considerable promise for resolving the structure of
the broad emission line region in quasars.

\end{abstract}


\section{Introduction}

Microlensing in the images of a multiply-imaged quasar was first
reported by \markcite{iwh+89}{Irwin} {et~al.} (1989) for the quadruple lens Q2237+0305
\markcite{huchra85}({Huchra} {et~al.} 1985).  Most quasar microlensing studies have been based on
broad-band photometric monitoring \markcite{wus+00,sus+03,cs03}(e.g., {Wo{\' z}niak} {et~al.} 2000; {Schechter} {et~al.} 2003; {Colley} \& {Schild} 2003),
which is sensitive primarily to variations in the continuum.
Microlensing of the continuum is expected since the optical/UV
continuum emission is thought to originate in a region that is
comparable in size to the Einstein radius of a typical star in a
typical lens galaxy.

Microlensing of the broad emission line region (BELR) is also
possible, if the BELR has structure on scales comparable to the
Einstein radius of a star \markcite{nem88,sw90}({Nemiroff} 1988; {Schneider} \& {Wambsganss} 1990).  The possibility of BELR
microlensing seemed rather remote until recent reverberation mapping
work revised the estimate of the BELR size downward from
$\sim\!10^{18}$ cm to $\sim\!10^{16}$ cm \markcite{wpm99,ksn+00}({Wandel}, {Peterson}, \& {Malkan} 1999; {Kaspi} {et~al.} 2000).
Inspired by these numbers, \markcite{amm+02}{Abajas} {et~al.} (2002) and \markcite{la03}{Lewis} \& {Ibata} (2003) revived the
idea of looking for microlensing of the BELR and computed possible
line profile variations for various BELR models.

Possible examples of microlensing of a quasar emission line have been
presented by \markcite{fil89}{Filippenko} (1989) for Q2237+0305, and by \markcite{cea+04}{Chartas et al.} (2003) for
H1413+117.  In particular, \markcite{cea+04}{Chartas et al.} (2003) detected a strong, redshifted
Fe K$\alpha$ emission line in the X-ray spectrum of only one of the
components in the quadruple lens H1413+117.  Although they did not have
multiple epochs to look for variability, \markcite{cea+04}{Chartas et al.} (2003) invoked a short
predicted time delay between the components to argue that microlensing
is the preferred explanation for seeing the Fe K$\alpha$ line in only
one component.

In this paper we present results from spectroscopic monitoring of the
recently discovered quadruple lens SDSS~J1004+4112
\markcite{inada03,oguri03}({Inada et al.} 2003; {Oguri et al.} 2003).  We observe variability in the broad emission
line profiles of one of the lensed images that provides strong
evidence for microlensing of the BELR, suggesting that the theoretical
predictions for microlensing were correct and confirming that the BELR
has structure on the scale of the Einstein radius of a star.

\section{The Data and Spectral Analysis}

Component B of SDSS~J1004+4112 was first identified as a quasar by
\markcite{cwh99}{Cao}, {Wei}, \& {Hu} (1999) and was also targeted as a quasar candidate
\markcite{rfn+02}({Richards} {et~al.} 2002) as part of the Sloan Digital Sky Survey (SDSS;
\markcite{yaa+00}{York} {et~al.} 2000); the SDSS spectrum was taken on 2003 February 3 and
the object was identified as a $z=1.734$ quasar.  Component A was also
identified as a possible quasar candidate based on its colors in the
SDSS imaging data \markcite{fig+96,gcr+98}({Fukugita} {et~al.} 1996; {Gunn} {et~al.} 1998); it was first confirmed as a
quasar (also at $z=1.734$) from observations taken on 2003 May 3 using
the ARC 3.5m telescope at Apache Point Observatory.  Higher quality
spectra of each of the four components and the lens galaxy were taken
on 2003 May 31 with the LRIS \markcite{occ+95}({Oke} {et~al.} 1995) spectrograph on the Keck I
telescope at the W.~M.~Keck Observatory.  These spectra are further
described in \markcite{inada03}{Inada et al.} (2003), \markcite{oguri03}{Oguri et al.} (2003), and
Table~\ref{tab:tab1}.  Table~\ref{tab:tab1} gives the UT date of the
observations, the telescope used, the components observed, the
exposure time, the spectral dispersion, and the wavelength range
covered.

\begin{figure}[p]
\epsscale{1.0}
\plotone{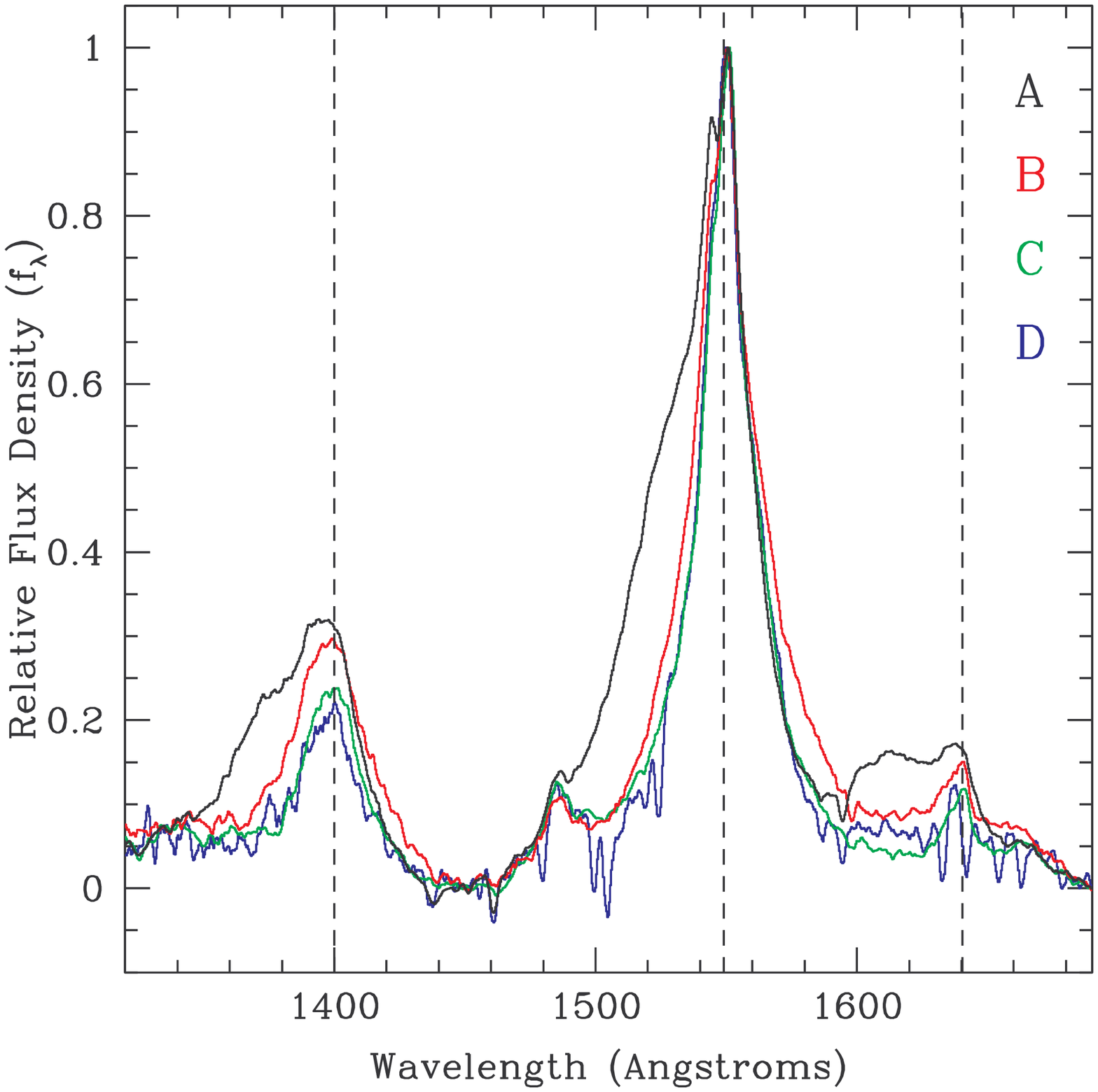}
\caption{Keck/LRIS spectra of components A, B, C, and D of
SDSS~J1004+4112.  Dashed vertical lines indicate the expected peaks of
\ion{Si}{4}/\ion{O}{4}], \ion{C}{4}, and \ion{He}{2} for $z=1.734$.
A power law continuum, fit between $\lambda\,1450$~\AA\ and
$\lambda\,1690$~\AA, has been subtracted from each spectrum, and the
spectra are all normalized to the peak of \ion{C}{4}.  The spectra
are smoothed by a seven pixel boxcar filter.
\label{fig:fig1}}
\end{figure}

\markcite{inada03}{Inada et al.} (2003) and \markcite{oguri03}{Oguri et al.} (2003) noted that the Keck/LRIS spectra
showed that all four components had the same redshift and similar
spectra, but there are some obvious differences --- with component A
showing the largest differences relative to the other components (see
Fig.~\ref{fig:fig1}).  Specifically, component A has a much stronger
blue emission line wing in the high-ionization lines
(\ion{Si}{4}/\ion{O}{4}], \ion{C}{4}, and \ion{He}{2}).  Component B
appears to have a slightly enhanced red wing as compared with the
other three components.  In Figure~\ref{fig:fig1}, we have sought to
emphasize the differences in the \ion{C}{4} emission line profile by
subtracting a continuum fit between $\lambda\,1450$~\AA\ and
$\lambda\,1690$~\AA, and then normalizing to the peak of \ion{C}{4}
emission.  Other choices in presentation could enhance or reduce the
appearance of similarities and differences between the components.

\begin{figure}[p]
\epsscale{1.0}
\plotone{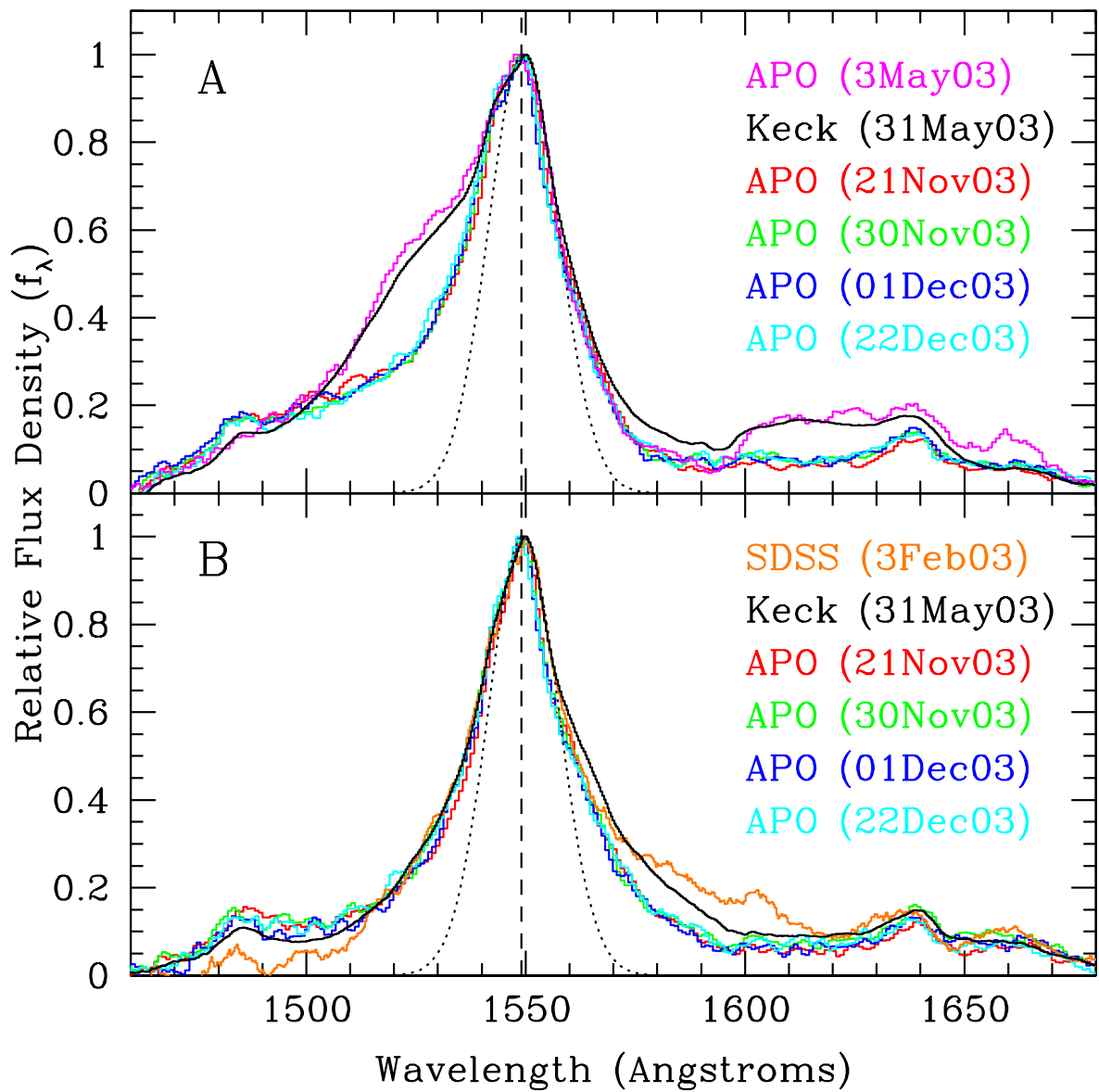}
\caption{Seven epochs of data in the \ion{C}{4} emission line region
of SDSS~J1004+4112.  A power law continuum, fit between
$\lambda\,1450$~\AA\ and $\lambda\,1690$~\AA, has been subtracted from
each spectrum, and the spectra are all normalized to the peak of
\ion{C}{4}.  We also overplot a scaled Gaussian at the center of
\ion{C}{4} to guide the eye toward emission line differences that are
persistent with time.  The dashed vertical line indicates the expected
\ion{C}{4} peak at $\lambda\,1549.06$~\AA.  All spectra have been
smoothed to similar resolutions.  Note that this smoothing hides the
associated absorption system that is observed just blueward of the
\ion{C}{4} emission line peak.
\label{fig:fig2}}
\end{figure}

Based on the puzzling differences in the broad emission line profiles
of the four components of the Keck/LRIS spectra, in 2003 November and
2003 December we obtained additional spectra of images A and B to
monitor the differences; see Table~\ref{tab:tab1}.  We used the
DIS\,III spectrograph on the ARC 3.5m telescope at Apache Point
Observatory, using the same instrument setup and data reduction for
each epoch.  The dispersion was $2.4\,{\rm \AA\,pixel}^{-1}$ and the
spectra cover the range $\lambda\lambda 3890$--$9350$~\AA.  Flux
calibration was performed with respect to a hot standard star: either
Feige~34 or G~191-B2B.  Wavelength calibration was performed with
respect to a helium-neon-argon comparison lamp.  Both components
(which are separated by $3\farcs73$) were observed at the same time in
order to minimize any differences between them, at the expense of loss
of spectrophotometric accuracy (slit not positioned at the parallactic
angle).  At each epoch we took two spectra of either 40 or 45 minutes
each.  These spectra were extracted separately and then combined to
yield 2003 November/December epoch spectra shown in
Figure~\ref{fig:fig2}.  The 2D spectra are only moderately resolved,
so we deblended the 2D spectra by fitting a double Gaussian profile
before extracting the spectra.  We estimate that contamination of the
faint component by the brighter component is less than 3\%.

Figure~\ref{fig:fig2} shows all seven epochs of spectroscopic data for
components A and B in the \ion{C}{4} emission line region.  The 2003
May 3 epoch observation of component A at APO confirms the reality of
the excess emission in the blue wing of component A in the Keck
spectrum.  The excess emission in \ion{He}{2} spans a velocity range
$\sim$2500--8500~km~s$^{-1}$ (blueshifted).  Both \ion{Si}{4} and
\ion{C}{4} have excesses over a slightly higher velocity range.  The
excess in \ion{He}{2} is much stronger in terms of equivalent width
than that in \ion{C}{4}.

\section{Discussion}

\subsection{Time-dependent Blue Wing Differences: Microlensing of the BELR}

\subsubsection{Discounting Alternatives and Objections}

To argue that the excess in the blue wings of the high-ionization
emission lines of component A is caused by microlensing of the BELR,
we must rule out alternative explanations.  The first question is
whether the flux enhancements are an artifact of the data acquisition
or reduction procedures.  We view this explanation as highly unlikely
because the enhancement appeared at two different epochs, in spectra
taken by different observers with different telescopes, and reduced
independently.

The second question is whether the spectral variations could be
intrinsic to the quasar, rather than induced by lensing.  If so, then
the same variations should be seen in the other lensed images, offset
in time by the lens time delays.  While the time delays are still
uncertain, in nearly all of the lens models presented by
\markcite{oguri03}{Oguri et al.} (2003) the delay between components A and B is predicted to
be $\lesssim$30 days.  (The Oguri et al.\ models do not form an
exhaustive set, but the prediction of a short time delay between A and
B is generic.) In $\sim$10\% of the models component A leads component
B.  In this case seeing the flux enhancement in the 2003 May 3 and
2003 May 31 spectra of component A but not in the 2003 May 31 spectrum
of component B means that intrinsic variability cannot explain the
data.  In the other $\sim$90\% of the models component B leads
component A.  In this case, the intrinsic variability hypothesis would
imply that the variations must have appeared in component B sometime
after 2003 February 3, been present during 2003 April (in advance of
the variations observed in component A in 2003 May), and then
disappeared before 2003 May 31.  In other words, the event must have
lasted $<$117 days in the observed frame, or $<$43 days in the rest
frame --- and we were fortunate to catch the event in component A just
before it disappeared.

Since the time separation of our spectral coverage does not completely
rule out intrinsic variability, this issue deserves further
discussion.  Specifically, we must address the likelihood of intrinsic
variability of the BELR on timescales of less than 43 days in the rest
frame.  Reverberation timescales can be less than 43 days since the
radius of the BELR is likely to be on the order of or less than this
size.  However, in that case the entire emission line profile will
vary with respect to any significant change in the continuum.  Since
we do not observe a significant change in the continuum level (see
below) and since the enhancement of the emission lines is only in the
blue wing and not over the entire profile, a reverberation effect is
unlikely.  Thus we are left with the possibility of a dynamical change
--- such as is seen in the so-called double-peaked emission line
quasars \markcite{eh03,ssh+03}(e.g., {Eracleous} \& {Halpern} 2003; {Strateva} {et~al.} 2003) --- which could better explain
the blue-wing-only nature of the enhancement in the high-ionization
emission lines.  Indeed the dynamical timescale ($\sim\!6$\,months;
\markcite{era04}{Eracleous} 2003) can be in the range needed to explain our
observations.  However, double-peaked emission is typically only seen
in the Balmer lines (and sometimes \ion{Mg}{2}), and is absent or weak
at best in high-ionization lines like \ion{C}{4} and \ion{He}{2}.
Furthermore, it would be very unusual for a double-peaked emission
line object to show a blue-wing enhancement in both our 2003 May 3 and
2003 May 31 epochs, but no strong blue- or red-wing enhancement of our
four 2003 November/December epochs.  Thus, although an explanation of
the data in terms of intrinsic variability is not rigorously excluded,
such a model would have to be rather improbably contrived.

One possible objection to the microlensing hypothesis is if the BELR
was microlensed, why not the continuum as well?  It is difficult to
put constraints on any enhancement of the continuum of component A
relative to B during the time spanned by our observations, because not
all of the spectra were taken at the parallactic angle.  Still, we can
estimate that component A was no more than $\sim$20\% brighter than B
in the continuum in 2003 May (as compared to 2003 November/December).
This is not obviously inconsistent with the microlensing hypothesis,
however.  It is easy to imagine configurations in which part of the
BELR is close enough to the caustic in the source plane of a star in
the lens to be microlensed, while the continuum source was far enough
to feel little effect.  In fact, this picture is consistent with our
hypothesis that only {\em part} of the BELR was being microlensed in
2003 May (see \S~\ref{sec:belr}).

\begin{figure}[p]
\epsscale{0.5}
\plotone{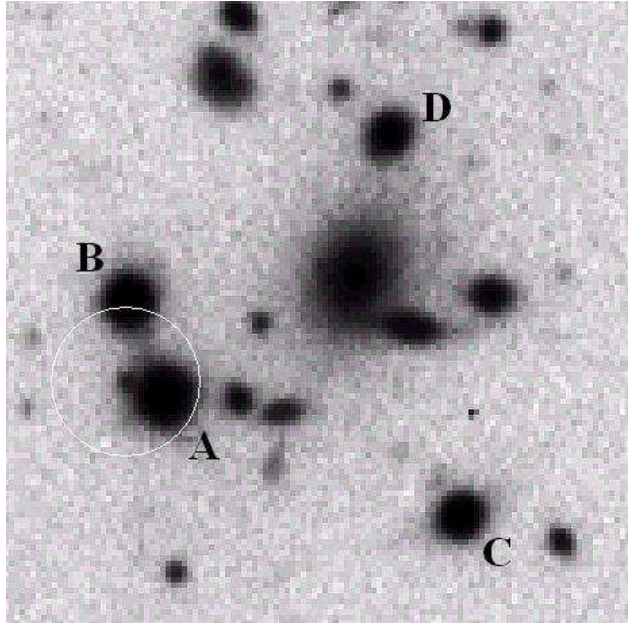}
\caption{Central ($22\farcs6$) region of the 1340\,s $i$-band Subaru
Prime Focus Camera image from \markcite{oguri03}{Oguri et al.} (2003, Fig.~8).  North is up,
East is left.  Although the image is saturated and proper PSF
subtraction of component A is not possible, there appears to be a
superimposed galaxy (with an estimated $i\sim24.5$) just to the
North-East of component A that could host the microlensing object.
The white circle is centered on the possible microlensing galaxy and
represents a $20\,h_{70}^{-1}\,{\rm kpc}$ radius at $z=0.68$ assuming
$\Omega_M=0.3$, $\Omega_\Lambda=0.7$, and $H_0=70\,h_{70}\,{\rm
km\,s^{-1}\,Mpc^{-1}}$.
\label{fig:fig3}}
\end{figure}

Another possible complication with the microlensing hypothesis is that
cluster galaxies tend to be stripped of their halos through mergers
and interactions.  Microlensing would therefore require either a
cluster member very close to the line of sight to component A or a
population of intracluster stars or massive compact halo objects
(MACHOs; \markcite{tot03,blz+04}e.g., {Totani} 2003; {Baltz et al.} 2003).  The presence of
intracluster MACHOs might not be surprising, because tidal forces
would naturally strip MACHOs from cluster galaxies along with the
galaxy halos.  On the other hand, there is evidence for a galaxy
superimposed on component A that could host the microlensing object,
as shown in Figure~\ref{fig:fig3}.  Proper PSF subtraction to confirm
this hypothesis is not possible since component A is saturated in this
image, but our best efforts do reveal residual flux at the center of
the circle in Figure~\ref{fig:fig3} with an estimated magnitude of
$i\sim24.5$.

\subsubsection{Examining the Microlensing Hypothesis}

We can also reverse the argument and ask if our microlensing
hypothesis makes sense given our current understanding of the
structure of quasars and the details of microlensing.  Without
invoking detailed microlensing and BELR models (which would be
premature since we have observed only one microlensing event), we
still find that qualitative and quantitative arguments reveal good
consistency between the microlensing hypothesis and the data.

The scale for microlensing is given by the Einstein radius of a star
\markcite{sef92}({Schneider}, {Ehlers}, \& {Falco} 1992),
\begin{eqnarray}
  \ReinS &=& \left(4\frac{GM}{c^2}\frac{D_{s} D_{ls}}{D_{l}}\right)^{1/2} \\
  \ReinL &=& \frac{D_{l}}{D_{s}}\,\ReinS
\end{eqnarray}
where $\ReinL$ and $\ReinS$ are the Einstein radius projected into the
lens and source planes, respectively, and $D_{l}$, $D_{s}$, and
$D_{ls}$ are the angular diameter distances to the lens, to the
source, and from the lens to the source, respectively.  Microlensing
is said to occur at low optical depth if the mean separation between
stars is $d \gg \ReinL$ and stars contribute a small fraction of the
surface mass density, or at high optical depth if $d \sim \ReinL$ and
stars contribute a substantial fraction of the surface mass density.
The microlensing probability can be high even if the optical depth is
low \markcite{sw02}({Schechter} \& {Wambsganss} 2002).  Regardless of whether the optical depth is high or
low, the Einstein radius sets the characteristic scale, and the BELR
can be microlensed if it has structure on scales $\lesssim\!\ReinS$
\markcite{amm+02,la03}({Abajas} {et~al.} 2002; {Lewis} \& {Ibata} 2003).  For SDSS~J1004+4112, the lens redshift is
$z_l=0.68$ and the source redshift is $z_s=1.734$, so the Einstein
radius of a $0.1\,M_{\sun}$ star associated with the lensing cluster
is
\begin{equation}
  \ReinS
  \sim 1.4\times10^{16} \left(\frac{M}{0.1M_{\sun}}\right)^{1/2}
    h_{70}^{-1/2} \mbox{ cm}
  \sim 5.3 \left(\frac{M}{0.1M_{\sun}}\right)^{1/2}
    h_{70}^{-1/2} \mbox{ lt-days}
\end{equation}
in a cosmology with $\Omega_M=0.3$, $\Omega_\Lambda=0.7$, and
$H_0=70\,h_{70}\,{\rm km\,s^{-1}\,Mpc^{-1}}$.  Thus, the typical
stellar Einstein radius is indeed comparable to the currently favored
size of the BELR, and microlensing is not unexpected.

The variability is caused by relative motion between the caustic
network and the source.  If the motion is dominated by the proper
motion of the lens galaxy (with transverse velocity $v_{\perp}$), then
the effective transverse velocity projected into the source plane and
expressed in distance per unit observed-frame time is
\begin{equation}
  v_{\rm eff} = \frac{v_{\perp}}{1+z_{l}}\,\frac{D_{s}}{D_{l}}\ .
\end{equation}
If the microlensing is dominated by caustic crossings, then the
characteristic event duration is the time for the caustic to sweep
across the source, $t_{\rm src} \sim 2 R_{\rm src} / v_{\rm eff}$.  If
we estimate $v_{\perp} \sim \sigma$ where $\sigma$ is the velocity
dispersion of the lens, then $\sigma \sim 700$~km~s$^{-1}$ for the
lensing cluster in SDSS~J1004+4112 \markcite{oguri03}({Oguri et al.} 2003) yields
\begin{equation}
  t_{\rm src} \sim 12.7 \left(\frac{R_{\rm src}}{10^{16}\mbox{ cm}}\right)
    \left(\frac{v_{\perp}}{700\mbox{ km s}^{-1}}\right)^{-1}
    \mbox{ yr}
\end{equation}
The fact that we see variability in the broad emission lines on a time
scale of $\sim$6 months suggests either that we happened to catch the
end of a long event, or that microlensing is affecting only part of
the BELR with a characteristic size smaller than $10^{16}$~cm (or
both).

A second interesting time scale is the typical time between
microlensing events.  Here the key unit is the time to cross an
Einstein radius,
\begin{eqnarray}
  t_{\rm Ein} &\sim& \ReinS / v_{\rm eff} \;\; \sim \; 8.6 \left(\frac{M}{0.1M_{\sun}}\right)^{1/2}
    \left(\frac{v_{\perp}}{700\mbox{ km s}^{-1}}\right)^{-1}
    h_{70}^{-1/2}\mbox{ yr}
\end{eqnarray}
If even a few percent of the mass is in stars, microlensing is
associated with a caustic network rather than a single star, and so
naive estimates of the time between events are difficult.  To obtain a
better estimate, we have used the standard ray shooting technique to
compute microlensing magnification maps and generate sample light
curves \markcite{krs86,wpk90}({Kayser}, {Refsdal}, \& {Stabell} 1986; {Wambsganss}, {Paczynski}, \&  {Katz} 1990).  We find that starting from a
non-microlensed position, the average wait time until the
magnification changes by $\ge$30\% is $\sim\!(0.2\mbox{--}0.8) \times
t_{\rm Ein}$.  The range represents uncertainties in the source size
and in the number density of microlenses.  Although there are many
uncertainties, it seems reasonable to expect that microlensing events
could be observed in SDSS~J1004+4112 on an approximately yearly basis.
Incidentally, if $t_{\rm src} \gtrsim t_{\rm Ein}$ then microlensing
events will blur together and the light curves will show continuous
smooth variations \markcite{koc04}(see, e.g., {Kochanek} 2003).  The apparent lack of
microlensing in the 2003 November/December epochs therefore adds
further support to the hypothesis that the region being microlensed is
smaller than $\sim\!10^{16}$~cm.

\begin{figure}[p]
\epsscale{1.0}
\plotone{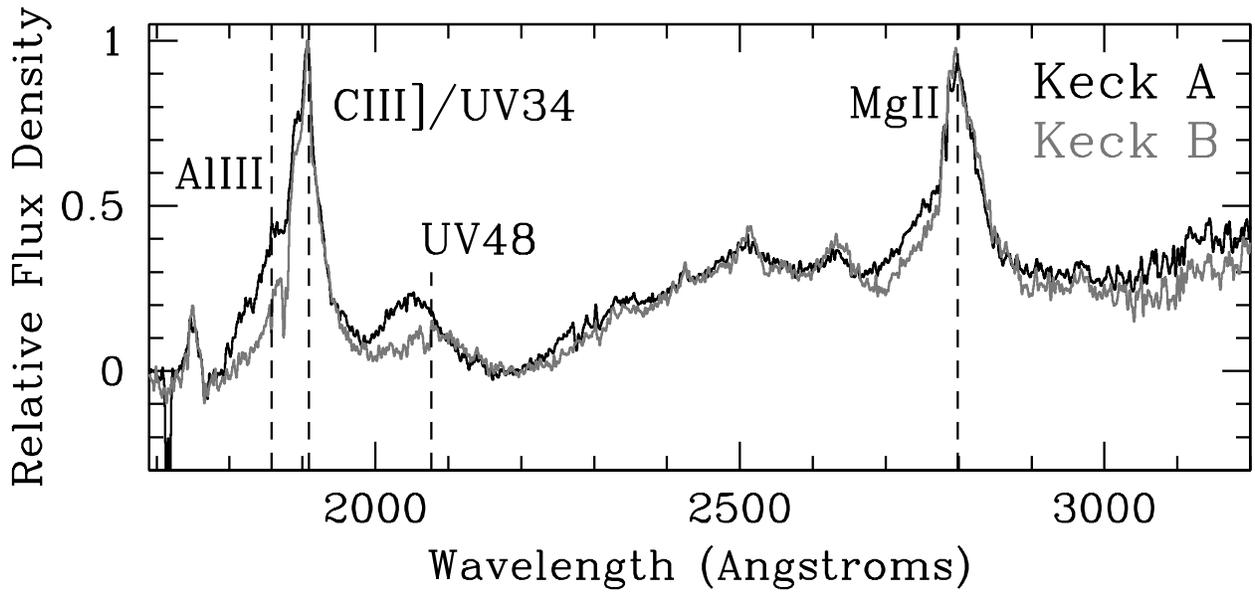}
\caption{Comparison of Keck spectra of components A and B in the
\ion{C}{3}] to \ion{Mg}{2} region of the spectra, showing the weaker
enhancement (relative to the higher ionization lines) of the blue
wings of these emission lines.  A power law continuum, fit between
$\lambda\,1690$~\AA\ and $\lambda\,2200$~\AA, has been subtracted
from each spectrum, and the spectra are all normalized to the peak of
\ion{C}{3}].  No constraint is placed on \ion{Mg}{2}.  The dashed
vertical lines indicate the expected \ion{Al}{3}, \ion{C}{3}] and
\ion{Mg}{2} peaks in addition to the \ion{Fe}{3} UV34 and UV48
complexes.  The spectra are smoothed by a seven pixel boxcar filter.
\label{fig:fig4}}
\end{figure}

The microlensing hypothesis nicely explains one of the more
interesting observational results, namely that different emission
lines show different amounts of variability.  From reverberation
mapping results, we know that the BELR is stratified by ionization and
that higher ionization lines are found closer to the center
\markcite{pw99}({Peterson} \& {Wandel} 1999).  As a result, the highest ionization regions have the
smallest effective sizes and should be the most sensitive to
microlensing.  Our finding that the excess in the \ion{He}{2} line is
stronger than the excess in the \ion{C}{4} line (in terms of
equivalent width) is consistent with their relative reverberation
mapping sizes \markcite{pw99}({Peterson} \& {Wandel} 1999).  Furthermore, there is a suggestion of a
weaker excess in the blue wings of the
\ion{C}{3}]/\ion{Si}{3}]/\ion{Al}{3}/FeIII\,UV34 complex, \ion{Mg}{2},
and \ion{Fe}{3} UV48 (a triplet at $\lambda\lambda\lambda$ 2062.211,
2068.904, 2079.652); see Figure~\ref{fig:fig4}, and also the ratio
spectra in Figure~3 of \markcite{oguri03}{Oguri et al.} (2003).  All the above lines are lower
ionization lines than \ion{C}{4} and \ion{He}{2}, so their weaker
excesses are also consistent with being more weakly microlensed as a
result of their emitting regions being larger with respect to the
projected Eistein radius of the lens.

An obvious question is why should microlensing of the BELR be seen in
this lens system, but not in others?  One possibility has to do with
the quasar's intrinsic luminosity.  The observed absolute $i$
magnitude of component A is $M_i=-26.9$, but this component is
amplified by a factor of $\sim$20 or more,\footnote{An isothermal
ellipsoid plus shear lens model, which is simple but fits the data
well, implies an amplification factor of 23 for component A.  More
complicated models yield a broad range of amplifications where the
median is 18 but there is a long tail to amplifications of 100 or
more.  See \markcite{oguri03}{Oguri et al.} (2003) for modeling details.} so the intrinsic
absolute $i$ magnitude is $M_i \gtrsim -23.8$.  Many other lensed
quasars have similar observed magnitudes but smaller amplifications,
and hence higher luminosities.  Since the size of the BELR scales as
the 0.5--0.7 power of the luminosity \markcite{ksn+00}(e.g., {Kaspi} {et~al.} 2000),
SDSS~J1004+4112 being relatively under-luminous would make it more
sensitive to microlensing than many other lensed quasars.

The large amplification helps make microlensing unsurprising in
another way.  Large amplifications are associated with significant
distortions that increase the size of the caustics (see Fig.~2 of
\markcite{sw02}{Schechter} \& {Wambsganss} 2002).  Thus, with a given star field the microlensing
probability increases as the amplification increases.  This may
explain not only why microlensing has been detected in
SDSS~J1004+4112, but also why it was seen first in component A (the
highest-amplification component).

\subsubsection{Implications for BELR Structure\label{sec:belr}}

Obtaining concrete constraints on the structure of the BELR will
require detailed modeling, and would greatly benefit from observations
of additional microlensing events.  Nevertheless, combining the data
in hand with general arguments already permits some strong and
valuable conclusions.  First, unless the number of clouds is orders of
magnitude smaller than required by other high-resolution Keck
observations \markcite{abl98}({Arav} {et~al.} 1998), the asymmetric nature of the microlensing
(lensing of the blue wing only) rules out a spherically symmetric
distribution of dynamically virialized, thermal line-width clouds
\markcite{nem88,amm+02,la03}({Nemiroff} 1988; {Abajas} {et~al.} 2002; {Lewis} \& {Ibata} 2003).

Pure radial outflow models can be ruled out because they would produce
symmetric line profile changes with most of the variation in the line
core \markcite{nem88}({Nemiroff} 1988).  Pure radial inflow models cannot be fully
excluded, but asymmetric microlensing would require the accretion disk
to have a radial extent comparable to the radial extent of the BELR
and to be optically thick (such that we do not see clouds at all
velocities).

The asymmetry therefore seems to imply that a strong rotational
component is needed in the high-ionization region of the BELR.  Such a
component could come in the form of a pure Keplerian disk or a
rotating disk-wind \markcite{mc98,elv00}(e.g., {Murray} \& {Chiang} 1998; {Elvis} 2000).  Microlensing of the
part of the BELR that is rotating toward us would then easily explain
the features that we observe.

The most robust statement we can make is that the observations confirm
that the BELR has structure on the scale of the Einstein radius.
Because we are sampling a region on the order of
$\sim\!1.4\times10^{16}$~cm, we effectively have a ``telescope'' with
a resolution of $\sim\!5\times10^{-7}$~arcseconds (given an angular
diameter distance to the quasar of $\sim5.4\times10^{27}$~cm).  In
other words, in SDSS~J1004+4112, nature has provided us with an
extremely powerful tool for the study of BELR structure.

\subsubsection{Some Predictions}

The microlensing hypothesis leads to several predictions that may
guide further observations.  First, the nature of the object(s)
responsible for the microlensing is unknown.  Two obvious
possibilities are stars or MACHOs associated with a galaxy (presumably
a cluster member) with a small projected impact parameters to
component A, or stars or MACHOs in the intracluster medium.  While the
Subaru $i$-band image of the field (Fig.~\ref{fig:fig3}) suggests that
there is indeed a galaxy near component A, the saturation of the
quasar image makes PSF subtraction uncertain.  Deep, high-resolution
images, preferably in the near-IR, would better reveal whether this
galaxy is real.  If so, then the microlensing optical depth would be
relatively high for component A and (perhaps quite) low for the other
components, so further monitoring would probably reveal additional
microlensing events in component A but not in the other components.

Alternatively, if the microlensing is caused by intracluster stars
or MACHOs then the microlensing optical depth for component A is
probably fairly low --- but it is likely to be similar for the other
components.  In this case further monitoring could well reveal
microlensing events in the other components, and the frequency of
events would reveal the number density of intracluster objects.

If additional microlensing events are detected, it will be extremely
interesting to track which lines are microlensed as a function of
time.  That would provide a unique and powerful new probe of
ionization stratification in the BELR.  In addition, the velocity
dependence of any enhancements of the BELR in future microlensing
events will help to reveal the kinematic structure of the BELR.

\subsection{Time-independent Emission Line Differences}

In addition to the time-dependent differences in the blue wings of the
high-ionization emission lines, there are also subtle differences in
the \ion{C}{4} emission line profiles between the components that
persist over at least 322 days in the observed frame.  In
Figure~\ref{fig:fig2} we have overplotted a Gaussian at the position
of \ion{C}{4} emission to guide the eye and help illustrate these
differences.

In particular, we note that there is a slight excess of
high-velocity redshifted \ion{C}{4} emission in component B as
compared with component A.  This excess is best seen in the 2003
May 31 Keck spectra, where there is a kink in the profile of B
near $\lambda\,1560$~\AA\ that is not present in component A.  In
addition, in all observations of component B the fall-off in the
red wing toward $\lambda\,1600$~\AA\ is more gradual than in
component A.  We also see similar (but opposite) differences in the
blue wing with the most extreme blue wing flux falling off more
gradually in component A than in component B.  In other words, at
all epochs the \ion{C}{4} emission line profile is somewhat
blueward asymmetric in component A, and somewhat redward asymmetric
in component B.

The cause of these differences is unclear.  Because the predicted A--B
time delay is $\lesssim$30 days, they are unlikely to be due to
intrinsic variability (unless the time delays are grossly
underestimated).  One interesting possibility is that they are also
due to microlensing, but with a mass scale that is much larger than
for the blue-wing BELR microlensing discussed above (in order to make
the variability time scale longer than 322 days in the observed
frame).  The responsible objects could perhaps be globular clusters,
or clumps of dark matter of mass $\sim\!10^4$--$10^8\,M_{\sun}$, in
which case the phenomenon would be termed millilensing (rather than
microlensing) and could provide a unique probe of dark matter
substructure of the type predicted in the Cold Dark Matter paradigm
\markcite{mm01,dk02,wbc+03,mmb+03}({Metcalf} \& {Madau} 2001; {Dalal} \& {Kochanek} 2002; {Wisotzki} {et~al.} 2003; {Metcalf et al.} 2003).

One way to test the millilensing hypothesis would be to normalize the
spectra using narrow lines since they should be insensitive to
small-scale structure \markcite{mm03}({Moustakas} \& {Metcalf} 2003).  Any differences in the broad
lines would then indicate millilensing.  This might be possible with
either J-band IR spectra of [\ion{O}{3}] $\lambda\lambda4959,5007$ or
with the unusually strong (but not apparently microlensed) nitrogen
lines seen in our optical spectra, since the nitrogen lines appear
relatively narrow and may lack a broad component.

\section{Conclusions}

We have presented seven epochs of spectroscopic data on the two
brightest components of the wide-separation, quadruply imaged quasar
SDSS J1004+4112.  Although the simplest lensing scenarios predict that
the four components should have identical spectra, the data reveal
significant differences in the emission line profiles of the
components.  In particular, the \ion{C}{4} emission line profile in
components A and B show both variable differences and differences that
are constant over 322 observed-frame days.  The \ion{He}{2} and
\ion{Si}{4}/\ion{O}{4}] lines in component A also show variability
similar to that seen in the \ion{C}{4} line.

Because the predicted time delay between A and B is $\lesssim$30 days,
we argue that the differences are not due to intrinsic variability in
the quasar coupled with a lensing time delay.  Instead, we suggest
that the variability in the blue wing of component A is best explained
by microlensing of part of the broad emission line region, resolving
BELR structure on the order of a few light days.  This represents the
first robust detection of BELR microlensing, with evidence based on
multiple emission lines and involving observed variability.  The
nature of the time-independent differences is less clear, but they may
also be the result of a lensing event.  In any case, it is clear that
continued spectroscopic monitoring of SDSS~J1004+4112 should be
carried out in an attempt to map the structure of its broad emission
line region through additional microlensing events.

\acknowledgements 

Funding for the creation and distribution of the SDSS Archive has been
provided by the Alfred P. Sloan Foundation, the Participating
Institutions, the National Aeronautics and Space Administration, the
National Science Foundation, the U.S.\ Department of Energy, the
Japanese Monbukagakusho, and the Max Planck Society.  The SDSS Web
site is http://www.sdss.org/.  The SDSS is managed by the
Astrophysical Research Consortium (ARC) for the Participating
Institutions. The Participating Institutions are The University of
Chicago, Fermilab, the Institute for Advanced Study, the Japan
Participation Group, The Johns Hopkins University, Los Alamos National
Laboratory, the Max-Planck-Institute for Astronomy (MPIA), the
Max-Planck-Institute for Astrophysics (MPA), New Mexico State
University, University of Pittsburgh, Princeton University, the United
States Naval Observatory, and the University of Washington.

Based on observations obtained with the Apache Point Observatory
3.5-meter telescope, which is owned and operated by the Astrophysical
Research Consortium.  Based on observations obtained at the W.~M. Keck
Observatory, which is operated as a scientific partnership among the
California Institute of Technology, the University of California and
the National Aeronautics and Space Administration, made possible by
the generous financial support of the W. M. Keck Foundation.  Based in
part on data collected at Subaru Telescope, which is operated by the
National Astronomical Observatory of Japan.  G.~T.~R. was supported in
part by HST grant HST-GO-09472.01-A.  C.~R.~K. is supported by NASA
through Hubble Fellowship grant HST-HF-01141.01-A from the Space
Telescope Science Institute, which is operated by the Association of
Universities for Research in Astronomy, Inc., under NASA contract
NAS5-26555.

\clearpage



\clearpage

\clearpage

\begin{deluxetable}{llllll}
\tabletypesize{\small}
\tablewidth{0pt}
\tablecaption{Summary of Observations\label{tab:tab1}}
\tablehead{
\colhead{Date (UT)} &
\colhead{Telescope} &
\colhead{Components} &
\colhead{Exp. Time} &
\colhead{Dispersion} &
\colhead{$\lambda$ Range} \\
\colhead{} &
\colhead{} &
\colhead{} &
\colhead{(sec)} &
\colhead{(${\rm \AA}\,{\rm pix}^{-1}$)} &
\colhead{(\AA)}
}
\startdata
2003 February 3 & SDSS 2.5m & B & 2700 & $\sim1$ & 3800-9200 \\
2003 May 3 & ARC 3.5m (DISIII) & A & 1800 & $\sim2.4$ & 3820-5630 \\
2003 May 31 & Keck I (LRIS) & ABCD & 900 & $\sim1$ & 3028-9700 \\
2003 November 21 & ARC 3.5m (DISIII) & AB & 2700+2700 & $2.4$ & 3890-9350 \\
2003 November 30 & ARC 3.5m (DISIII) & AB & 2700+2700 & $2.4$ & 3890-9350 \\
2003 December 1 & ARC 3.5m (DISIII) & AB & 2400+2400 & $2.4$ & 3890-9350 \\
2003 December 22 & ARC 3.5m (DISIII) & AB & 2400+2400 & $2.4$ & 3890-9350 \\
\enddata
\end{deluxetable}

\end{document}